\documentclass[aip,jap,preprint,superscriptaddress]{revtex4-2}
\usepackage{graphicx}
\usepackage{dcolumn}
\usepackage{bm}
\usepackage{graphicx,color}
\usepackage{epstopdf}
\usepackage{commath}
\usepackage{appendix}
\usepackage{ulem}
\usepackage{comment}

\begin{document}


\title{Voltage Control of Electromagnetic Properties in Antiferromagnetic Materials}

\author{Xinyi Xu}
\affiliation{Department of Electrical and Computer Engineering, North Carolina State University, Raleigh, NC 27695, USA}
\author{Yuriy G. Semenov}
\affiliation{Department of Electrical and Computer Engineering, North Carolina State University, Raleigh, NC 27695, USA}
\affiliation{V. Lashkaryov Institute of Semiconductor Physics,  National Academy of Sciences of Ukraine, Kyiv 03680, Ukraine}
\author{Ki Wook Kim}\email{kwk@ncsu.edu}
\affiliation{Department of Electrical and Computer Engineering, North Carolina State University, Raleigh, NC 27695, USA}
\affiliation{Department of Physics, North Carolina State University, Raleigh, NC 27695, USA}

\begin{abstract}
Dynamic modulation of electromagnetic responses is theoretically examined in dielectric antiferromagnets.  While both magneto-electric and magneto-elastic coupling can achieve robust electrical control of magnetic anisotropy, the latter is considered in a bilayer structure with a piezoelectric material.   Numerical calculations based on the frequency-dependent permeability tensor clearly illustrate that the anisotropy profile in the typical uniaxial or biaxial antiferromagnets such as NiO and Cr$_2$O$_3$ can be modified sufficiently to induce a shift in the resonance frequency by as much as tens of percent in the sub-mm wavelength range (thus, an electrically tunable bandwidth over 10's of GHz).  The polarization of the electromagnetic response is also affected due to the anisotropic nature of the effect, offering a possibility to encode the signal.  The intrinsic delay in switching may be minimized to the ns level by using a sufficiently thin antiferromagnet.  Application to specific devices such as a bandpass filter further illustrates the validity of the concept.

\end{abstract}
\maketitle

\section{Introduction}
Spintronics has emerged as a key research field in the pursuit of energy efficient devices for next-generation information processing and storage.~\cite{Spintronics2020}  In particular, recent studies in the literature  have revealed a significant promise of antiferromagnets (AFMs) in a wide range of applications.~\cite{Bai2020, Fukami2020}
One major advantage of the AFMs over the far more investigated ferromagnets (FMs) is the ultra-high resonance frequencies (typically in the 100’s of GHz and above), which can in turn enable ultra-fast magnetization dynamics, such as switching and oscillation, as well as applications beyond the logic/memory realm in the technologically challenging THz regime.  Furthermore, the energy required to modulate the AFM magnetic state can reach the aJ level, about two-three orders of magnitude smaller than the corresponding values in the FMs.  The AFMs are also attractive for their insensitivity to the external magnetic fields and the absence of stray or demagnetization field.  This implies that the information stored in an AFM is robust against the disturbing magnetic fields and would not be affected by its neighbors, independent of how densely they are integrated. These virtues of AFMs originate from the compensated nature of their magnetic moments. Instead of having all magnetic moments aligned, those on each atomic site or layer in an AFM alternate in their directions with zero (or nearly zero) net magnetization.  While AFMs are insensitive to the perturbation by external magnetic fields due to the symmetry as mentioned above, it has been shown that the AFM order parameter can be manipulated via electrical,~\cite{Gomonay2018} optical,~\cite{Nemec2018} or thermal~\cite{Seki2015} approaches.


Antiferromagnetic resonance (AFMR), as one of the more investigated electromagnetic (EM) mechanism, can be understood in a manner similar  to that of ferromagnetic resonance.~\cite{Sun2012,Kittel1951,Kepfer1952}  It is intuitively evident that the EM response depends on the magnetic properties of the materials such as exchange coupling and magnetic anisotropy.
The studies  in the literature have already illustrated this point in the FMs by exploiting electrical modulation of these properties
via the magneto-electric and/or -elastic effects.~\cite{Sun2012} The magnetic anisotropy at the interface between a thin magnetic film and a nonmagnetic material was found to depend sensitively on the electric field applied to the layered structure.~\cite{Maruyama2009,Amiri2012,Nozaki2016,Alzate2014,Williamson2017,Kang2017,Peng2018} Likewise, bilayer composites with piezoelectric materials have demonstrated effective control of magnetic properties by strain mediated coupling of the magnetic and electric order parameters.~\cite{Vaz2012,Gomonay2014} Considering the physical characteristics of these effects, essentially similar responses can be expected when the free layer is made of an AFM instead.~\cite{Semenov2017,Barra2018}

In this work, we theoretically explore active manipulation of EM responses in the sub-mm wave frequencies with a broadly tunable spectral range by utilizing electrical modulation of magnetic properties in AFMs.  Of the two mechanisms for anisotropy control discussed above, the magneto-elastic effect
is the primary subject of investigation given that the dimensions of the AFM layers involved (such as the layer thickness) are likely to be of the order of microns due to the relatively long wave lengths of EM waves.  Accordingly, the phenomenon relying on interface charges may not be as effective.
Combination with a piezoelectric layer such as PZT or PMN-PT was already shown to induce the desired strain-mediated changes in the anisotropy profiles of relatively thick FMs.~\cite{Sun2012} Following the derivation of magnetic permeability tensor for uniaxial and biaxial AFMs, our numerical calculation shows that the bias induced changes in the magnetic anisotropy can shift the resonance frequency (i.e., the pole in the magnetic susceptibility) significantly $-$ as much as tens of percent $-$ offering a tunable frequency window of  10's of GHz or even larger.  This impact on the resonant frequency (thus, the permeability tensor) appears  particularly drastic when the strength of the modified out-of-plane magnetic anisotropy approaches to that of the primary easy axis resulting in a biaxial configuration.  Application to a sample device structure $-$ a bandpass filter $-$ illustrates a large range of central resonant frequency shift (i.e., tunability) coupled with a narrow linewidth only weakly dependent of resonant frequency modulation, offering potentially an extensive operable bandwidth. The dependence on the incident direction and polarization of the EM signal is also analyzed.


\section{Theoretical Model}
The AFMR was first analyzed in 1951.~\cite{Kepfer1952,Kittel1951} The dynamics of a bipartite monodomain AFM can be described conveniently by the Landau-Lifshitz-Gilbert (LLG) equation applied to each sublattice magnetization $\mathbf{M}_{1}$ and $\mathbf{M}_{2}$; i.e.,
\begin{equation}
\frac{d\mathbf{M}_{n}}{dt}=- \gamma \mathbf{M}_{n}\times \mathbf{H}_{n}+\mathbf{R}_{n}; ~~~~ n=1,2,
\label{LLn}
\end{equation}
where the effective magnetic field and the dissipation term can be expressed, respectively, as
\begin{equation}
\mathbf{H}_{n}=- \frac{\delta W}{\delta \mathbf{M}_{n}}  \label{H12}
\end{equation}%
and
\begin{equation}
\mathbf{R}_{n}=\lambda \frac{\mathbf{M}_{n}}{M_{s}}\times \frac{d\mathbf{M}%
_{n}}{dt}.    \label{R12}
\end{equation}
Here,  $\gamma$ denotes the  gyromagnetic ratio, $M_s$ the sublattice saturation magnetization, and $\lambda$ the dimensionless damping parameter.  The contributions to the density of the magnetic energy $W$ considered in this work are attributed to the inter-sublattice exchange interaction ($\mathbf{H}_{n,ex}$) and the magnetic anisotropy ($\mathbf{H}_{n,an}$) along with a weak ac Zeeman field (i.e., an external ac field signal $\mathbf{h}e^{i\omega t}$). As the ground state of an AFM stipulates $\mathbf{M}_{2}=-\mathbf{M}_{1}$ in the absence of the Dzyaloshinskii–Moriya interaction,  the condition $\mathbf{H}_{2,ex}= - \mathbf{H}_{1,ex}$ naturally follows.  Similarly, the anisotropy field in a biaxial AFM (as shown in Fig.\ 1) also satisfies $\mathbf{H}_{2,an}= - \mathbf{H}_{1,an}$  as $\mathbf{H}_{n,an}$ is given approximately as $ ( \frac{K_{x}}{M_{s}^{2}}\mathbf{M}_{n,x}+ \frac{K_{z}}{M_{s}^{2}}\mathbf{M}_{n,z})$.  Here, $K_{x} $ ($K_{z} $) denotes the easy-axis anisotropy in the $x$ ($z$) axis, while $\mathbf{M}_{n,x}$ ($\mathbf{M}_{n,z}$) is the vector component of the sublattice magnetization along the corresponding direction.
In the following analysis, we consider either the uniaxial easy anisotropy along the $x$ (or $z$) axis or the biaxial anisotropy with both $x$ and $z$ as the easy directions. By applying a linear-order perturbation analysis in Eq.\ (\ref{LLn}), the desired susceptibility tensor can be obtained in the form of $\textbf{m} =\overleftrightarrow{\chi} \textbf{h}$, where $\textbf{m}=\textbf{m}_1+\textbf{m}_2$ and $\mathbf{m}_{n} e^{i\omega t} = \mathbf{M}_{n}-\mathbf{M}_{n 0}$ ($\mathbf{M}_{n 0}$ being the equilibrium state with $\mathbf{h}=0$).  While the properties of an AFM is unchanged once given, its response to an EM wave can vary dependent on the specifics of the incoming signal such as the polarization and the propagation direction.  A number of typical examples are described below.

When the $x$-directional easy anisotropy becomes dominant (i.e., $K_x > K_z $, $ K_x  > 0$), $\mathbf{M}_{n 0}$ lies essentially along this axis and the influence of the time varying perturbation becomes negligible (i.e., ${m}_{n,x} \approx 0$). A subsequent analysis shows that the corresponding susceptibility tensor is given as
\begin{equation}
\overleftrightarrow{\chi}^{[x]}=\left(
\begin{array}{ccc}
0 & 0 & 0 \\
0 & \chi _{yy}^{[x]} & 0 \\
0 & 0 & \chi _{zz}^{[x]}%
\end{array}%
\right) \label{permeability}
\end{equation}
with the diagonal elements
\begin{eqnarray}
\chi _{yy}^{[x]} & =& -\frac{2\gamma M_{s}(\gamma H_{A,x}-\gamma H_{A,z}+i\lambda \omega )}{\omega ^{2}-\omega _{r,y}^{2 [x]}-2i\lambda
\omega \gamma H_{E}},  \label{chi_yy} \\
\chi _{zz}^{[x]} &=& -\frac{2\gamma M_{s}(\gamma H_{A,x}+i\lambda \omega )}{\omega ^{2}-\omega _{r,z}^{2 [x]}-2i\lambda
\omega \gamma H_{E}}  \, .  \label{chi_zz}
\end{eqnarray}
When the real and imaginary parts are separated,  Eqs.\ (\ref{chi_yy}) and (\ref{chi_zz}) reduce to
\begin{eqnarray}
\chi _{yy}^{[x]} & =& -\frac{2\gamma^2 M_{s}(H_{A,x}-H_{A,z})(\omega ^{2}-\omega _{r,y}^{2 [x]})-4\gamma^2(\lambda\omega)^2 M_{s}H_{E}}{(\omega ^{2}-\omega _{r,y}^{2 [x]})^2+(2\lambda \omega \gamma H_{E})^2} \nonumber \\
&-& i\frac{2\gamma(\lambda \omega)M_{s}(\omega ^{2}-\omega _{r,y}^{2 [x]})-2\gamma^2 M_{s}(H_{A,x}-H_{A,z}) 2\gamma(\lambda\omega)H_{E}}{(\omega ^{2}-\omega _{r,y}^{2 [x]})^2+(2\lambda \omega \gamma H_{E})^2}  \label{chi_yy_c} \\
\chi _{zz}^{[x]} &=& -\frac{2\gamma^2 M_{s} H_{A,x}(\omega ^{2}-\omega _{r,z}^{2 [x]})-4\gamma^2(\lambda\omega)^2 M_{s}H_{E}}{(\omega ^{2}-\omega _{r,z}^{2 [x]})^2+(2\lambda \omega \gamma H_{E})^2}   \nonumber \\
&-& i\frac{2\gamma(\lambda \omega)M_{s}(\omega ^{2}-\omega _{r,z}^{2 [x]})-2\gamma^2 M_{s}H_{A,x} 2\gamma(\lambda\omega)H_{E}}{(\omega ^{2}-\omega _{r,z}^{2 [x]})^2+(2\lambda \omega \gamma H_{E})^2} \, .  \label{chi_zz_c}
\end{eqnarray}
The eigen-frequencies along the $y$ and $z$ axes are given as
\begin{eqnarray}
\omega_{r,y}^{2 [x]} &=&\gamma ^2 (H_{A,x}-H_{A,z})(2H_{E}+H_{A,x}) , \label{omega_ry} \\
\omega_{r,z}^{2 [x]} &=&\gamma ^2 H_{A,x}(2H_{E}+H_{A,x}-H_{A,z})  .  \label{omega_rz}
\end{eqnarray}
Thus, it is evident that a maximum response can be expected when the incoming EM wave is chosen to propagate in the same direction as the primary easy axis (i.e., non-zero $h_y$ and $h_z$).  The superscript $[x]$ in the above expressions is used to specify the primary easy axis, while $H_E$ ($= |\mathbf{H}_{ex}|$), $H_{A,x}$ ($ =\frac{ K_x}{M_s}$), and $H_{A,z}$ ($ = \frac{ K_z}{M_s}$) denote the strength of the exchange and normalized anisotropy fields, respectively. Note that in the case of $ |K_x|  \gg |K_z| $, both eigen-frequencies in Eqs.\ (\ref{omega_ry}) and (\ref{omega_rz}) can be approximated as $ \omega_r = \gamma\sqrt{2H_E H_A+H_A^2} $, converging to Kittle's equation for FMs.
The permeability tensor $\overleftrightarrow{\mu}^{[x]}$ is obtained subsequently as $\overleftrightarrow{I}+4 \pi \overleftrightarrow{\chi}^{[x]}$, where  $\overleftrightarrow{I}$ is the identity tensor.

The susceptibility tensor for the $z$-directional primary easy axis (i.e., $K_z > K_x $, $K_z  > 0$) can be described likewise with only the diagonal components $ \chi _{xx}^{[z]}$ and $ \chi _{yy}^{[z]}$.  The detailed expressions can be obtained by simply replacing the subscript $x$ to $z$ and vice versa in Eqs.\ (\ref{chi_yy})-(\ref{omega_rz}). A general form of the susceptibility for a biaxial AFM
can be derived from Eq.\ (\ref{permeability}) via a linear transformation of the coordinates ($x,y,z \rightarrow x^\prime,y^\prime,z^\prime$)
with the transformation matrix
\begin{equation}
U=\left(
\begin{array}{ccc}
\cos \theta \cos \varphi & \cos \theta \sin \varphi & \sin \theta \\
- \sin \varphi & \cos \varphi & 0 \\
- \sin \theta \cos \varphi  & - \sin \theta \sin \varphi & \cos \theta
\end{array}
\right) , \label{Umatirx}
\end{equation}
where the definition for angles $\theta$ and $\varphi$ can be found in Fig.\ 1.  Namely, a vector in the ($x$,$y$,$z$) coordinate can be translated into the $(x^\prime,y^\prime,z^\prime)$ system through the relation $\mathbf{r}^\prime = U \mathbf{r}$.
Then, the corresponding susceptibility tensor [with the easy axis now oriented along an arbitrary angle $\theta$,$\varphi$ in the ($x^\prime,y^\prime,z^\prime$) system] can be obtained from the relation $\overleftrightarrow{\chi} ^{[\theta,\varphi]}=U\overleftrightarrow{\chi} ^{[x]}U^{-1}$ as
\begin{equation}
\overleftrightarrow{\chi} ^{[\theta,\varphi]}=\left(
\begin{array}{ccc}
\chi _{yy}^{[x]}\cos ^{2}\theta \sin ^{2}\varphi +\chi _{zz}^{[x]}\sin
^{2}\theta  & \frac{1}{2} \chi _{yy}^{[x]} \cos \theta \sin 2\varphi
& \frac{1}{2} \sin 2\theta\ (\chi _{zz}^{[x]} - \chi _{yy}^{[x]}\sin^{2} \varphi)
\\
\frac{1}{2}\chi _{yy}^{[x]}\cos \theta \sin 2\varphi
 & \chi _{yy}^{[x]}\cos ^{2}\varphi  & -\frac{1}{2}\chi _{yy}^{[x]}\sin \theta\sin 2\varphi  \\
\frac{1}{2}\sin 2\theta (\chi _{zz}^{[x]}-\chi _{yy}^{[x]}\sin^{2} \varphi)  & -\frac{1}{2}\chi
_{yy}^{[x]}\sin \theta \sin 2\varphi  & \chi _{zz}^{[x]}\cos ^{2}\theta +
\chi _{yy}^{[x]} \sin ^{2}\theta \sin ^{2}\varphi
\end{array}%
\right) .
\end{equation}

The cases for different polarizations can be described by imposing appropriate conditions to the perturbation field $\mathbf{h}$. For instance, the characteristic response to circularly polarized waves around the easy $x$-axis can be obtained by setting $h_y=h$ and $h_z=h e^{\pm i\pi/2}$ for a $\pi/2$ phase shift.  Interestingly, the trajectories of $\mathbf{m}$ become elliptical when the material is not isotropic on the plane normal to the easy axis (e.g., $K_z \neq 0$).  This is apparent from Eqs.\ (\ref{permeability})-(\ref{chi_zz_c}) with $m_y$ and $m_z$ given by $ \chi_{yy}^{[x]} h$ and $ \chi_{zz}^{[x]} h e^{\pm i\pi/2}$, respectively.

As for electrical modulation of the magnetic anisotropy  (thus, $H_{A,x}$ and $H_{A,z}$),  the strain-induced effective anisotropy field mediated by the piezoelectric layer can be expressed in an empirical form as~\cite{Sun2012,Shepley2015}
\begin{equation}
\Delta H_A = \frac{3 \lambda_{s}}{M_s}{Y} \varepsilon  , \label{deltaH}
\end{equation}
where $\lambda_{s}$ and $Y$ denote the saturation magnetostriction constant and the Young's modulus of the magnetic material, respectively, and  $ \varepsilon$ is the strain applied in the direction of interest.  Considering the geometry of the bilayer structure with an electric field $E$ applied across the piezoelectric material (i.e., $z$; see Fig.\ 1), the strain induced in the in-plane direction (e.g., $x$) of the magnet and its counterpart in the normal direction ($z$) can be estimated approximately as $d_\mathrm{eff} E$ and $- \frac{2 \nu}{1- \nu} d_\mathrm{eff} E$, respectively, in a simple isotropic treatment.  Here, $d_\mathrm{eff} $ is the effective piezoelectric coefficient of the piezoelectric material and $\nu$ is the Poisson's ratio of the magnetic layer.  More precisely, $\lambda_s$, $d_\mathrm{eff}$, and  the ratio of the strain components are dependent on the crystallographic symmetry and orientation of the AFM/piezoelectric structure.  Nevertheless, these approximate expressions are sufficient to estimate the typical variable range of $\Delta H_A$ achievable in the $x$ and $z$ anisotropy fields (i.e., $\Delta H_{A,x}$ and $\Delta H_{A,z}$, respectively).~\cite{Sun2012,Shepley2015}  Both positive and negative values (i.e., the increase or decrease) are possible by switching  the polarity of the applied electric field (thus, the tensile or compressive strain).  As apparent from the differing signs, $\Delta H_{A,x}$ and $\Delta H_{A,z}$ shift in the opposite directions.

In the practical applications, the performance of an EM device or structure is often characterized by the scattering parameters as they can provide the information on key properties of the network such as the gain, loss, and impedance. Evidently, these quantities (frequently known as the $S$ parameters or $S$ matrix) are directly related to the permittivity and permeability of the material.  For instance, the $\{11\}$ component of this matrix for a dielectric waveguide is given as
\begin{equation}
    {S}_{11}=\frac{\Gamma(1-Z^2)}{1-\Gamma^2 Z^2}
\end{equation}
in terms of the reflection and transmission coefficients, $\Gamma=\frac{f(\mu)-1}{f(\mu)+1}$ and $Z=\mathrm{exp}(-\gamma L)$, respectively.  In turn,
$f(\mu)$ is defined by $f(\mu)=\mu\frac{\gamma_0}{\gamma}$, where $\gamma$ and $\gamma_0$ can be expressed in Gaussian units as
\begin{equation}
    \gamma=j\sqrt{\mu\epsilon\omega^2-(\frac{2\pi}{\lambda_c})^2} ,
\end{equation}.
\begin{equation}
    \gamma_0=j\sqrt{\omega^2-(\frac{2\pi}{\lambda_c})^2}
\end{equation}
via the permeability $\mu$,  permittivity $\epsilon$, sample length $L$, and cutoff wavelength $\lambda_c$.~\cite{Domich1991}
Modulation in the magnetic susceptibility (thus, the permeability) described above is clearly expected to manifest through the $S$ parameters.

\section{Results and Discussion}
A schematic of the bilayer structure consisting of magnetic and piezoelectric layers is shown in Fig.\ 1. As mentioned above, the strain-induced anisotropy energy provides an effective method for magnetic property modulation. Both the in-plane ($x$) and out-of-plane  ($z$) anisotropy are under control by an external electric bias or field which can apparently range over $\pm 10$ kV/cm   even in relatively thick samples.~\cite{Maruyama2009,Liu2009afm}  Unless specified otherwise, it is assumed that the incoming EM wave is propagating along the $x$ axis, which is also the direction of the initial easy axis. In addition, the frequencies $f(=\omega/2\pi)$ are used in the discussion instead of  $\omega$. For the numerical calculation, PMN-PT is adopted as the prototypical  piezoelectric layer with $d_{\mathrm{eff}}=-2000$ pC/N derived from $d_{\mathrm{eff}}=(d_{31}-d_{32})/(1-\nu_{p})$,  $d_{31}=-1800$ pC/N, $d_{32}=900$ pC/N, and Possion's ratio $\nu_{p}=0.32$.\cite{Liu2009,Jin2022} For the AFMs, NiO and Cr$_2$O$_3$ are chosen as two primary examples of the dielectric materials.  Due to the loss, metallic magnets are not preferred in general.

In the case of NiO, the magnetic anisotropy is characterized by a strong hard axis (say, $z$) and a weak easy axis ($x$); thus, essentially an easy-plane AFM.  Its magnetic and elastic properties can readily be found in the literature to have~\cite{Fraune2000,Gaillac2016,Hubert2015} $K_x = 6$ kJ/m$^3$, $K_z = -140$ kJ/m$^3$, $H_E= 382.5$ kOe, $M_s=350$ kA/m,
$\lambda_s= 30$ ppm (or $\mu\varepsilon$), $Y= 200$ GPa, and Poisson's ratio $0.27<\nu<0.32$.  For convenience, we use $\nu \approx 0.31$ which reduces the ratio  $\Delta H_{A,z}/\Delta H_{A,x}$ to a rounded number, $-0.9$.
These give the normalized anisotropy fields of the initial state (i.e., before strain) as  $H_{A,x}^0=340$ Oe and $H_{A,z}^0 = -80$ kOe. The strain-induced in-plane anisotropy change (i.e., $\Delta H_{A,x}$)  in NiO is approximately $ -103$ Oe$\cdot$cm/kV obtained from Eq.\ (\ref{deltaH}).

Figure 2(a) provides the calculated resonance frequencies $f_{r,y}$ ($={\omega_{r,y}}/ {2\pi}$; line 1) and  $f_{r,z}$ ($={\omega_{r,z}}/{2\pi}$; line 2) as a function of the applied electric field $E$.  The corresponding changes in $K_{x}$ and $K_{z}$  are also shown with $\Delta K_{z} / \Delta K_{x} = -0.9$ as specified above. The dissipation term $\lambda$ from Eq.\ (\ref{R12}) is neglected in this calculation for simplicity.  Due to a large negative value in $K_z$ at $E=0$, the range of the electric field under consideration (i.e., $\pm 10$ kV/cm) is insufficient to alter $z$ as the dominant hard axis (i.e., $K_z < 0$ throughout the range).  On the other hand, $K_{x}$ (thus, $H_{A,x}$)  becomes negative when $ E > +3.3$ kV/cm (marked by the vertical dashed line).  Note that this region is outside the consideration since the assumption of $x$ as the primary easy axis and thus $m_x \approx 0$ may no longer be valid. With the $E$-field induced strain,
$f_{r,y}$ can be modulated between $\sim$248 GHz and $\sim  $214 GHz from the initial value of $\sim $224 GHz.  The corresponding change in $f_{r,z}$ appears more drastic; i.e., from near zero to $\sim $91 GTHz ($\sim $45 GHz at $E=0$). The wide range of control with $\Delta f$/$f$ of over 10\% clearly illustrates the advantage of the approach under investigation. For a more comprehensive analysis, Fig.\ 2(b) shows both the $yy$ and $zz$ components of the permeability tensor for five different bias conditions; i.e.,  $ E=$ $0$, $\pm0.5$ kV/cm and $\pm1$ kV/cm.  Even for a relatively small electric field of $0.5$ kV/cm inducing $\Delta H_{A,x}$ of $50$ Oe, sizable frequency shifts of $\sim$2.6 GHz and $\sim$7 GHz  are observed in $\mu_{yy}^{[x]}$ and $\mu_{zz}^{[x]}$, respectively, making them clearly identifiable with a reasonably narrow linewidth.

Along with NiO, Cr$_2$O$_3$ provides another prototypical example. Unlike the former, this material does not exhibit a strong hard-axis anisotropy (say, along $z$), making it more amenable  for  externally controlled switching in the primary easy direction (thus, the magnetization orientation).  Moreover, its characteristic material properties indicate that Cr$_2$O$_3$ has a potential to be more responsive to the application of strain than NiO. The typical values found in the literature~\cite{Gaillac2016,Cr2O3,Mahmood2021} suggest $\lambda_s= 28$ ppm, $Y= 308$ GPa, $M_s=190$ kA/m, $K_x = 20$ kJ/m$^3$, $K_z =0$, and $H_E= 10$ MOe. These set the normalized initial anisotropy fields to $H_{A,x}^0=2.1$ kOe and $H_{A,z}^0= 0$ Oe.
While the Poisson's ratio has a slightly wider range than NiO with $0.25<\nu<0.35$, $\Delta H_{A,z}/\Delta H_{A,x}$ is set to $-0.9$ as before for convenience.  Since Cr$_2$O$_3$ has a uniaxial anisotropy (i.e., only non-zero $K_x$), the resonance frequencies satisfy $ f_{r,y}= f_{r,z}\sim 574$ GHz before the external modulation.

The strain-induced magnetic anisotropy $\Delta H_{A,x}$ is estimated to be $-270$ Oe$\cdot$cm/kV sufficient to result in an easy-axis flip (e.g., $x$$\rightarrow$$z$), unlike the NiO example, with an electric field up to $\pm 10$ kV/cm.  The subsequent reorientation in the sublattice magnetizations can take place relatively fast and energy efficiently in the AFMs.~\cite{Lopez2019,Peng2021}  While the intrinsic 90$^\circ$ switching may only require $\lesssim $10 ps,~\cite{LXL2017} the actual transient dynamics are likely to be dictated by the strain propagation through the structure (thus, the sound velocity and the thickness of the AFM film).~\cite{Barra2018}  The inset to Fig.\ 3(a) schematically illustrates the 90$^\circ$ rotation when the easy axis of the AFM layer changes from the in-plane to the out-of-plane $z$-direction.  Assuming a thickness of the order of 1 $\mu$m, this transition could typically be achieved in about 1 ns or so.
Here, a point to note is that non-zero components of the susceptibility tensor are also switched  correspondingly as discussed in Sec.\ II. This fact is clearly reflected in Fig.\ 3(a), where $f_{r,y}$ and $f_{r,z}$ (lines 1 and 2, respectively) are shown on the left side of the vertical line (i.e., easy $x$ dominant), and $f_{r,y}$ and $f_{r,x}$ (lines 1 and 3, respectively) on the right (i.e., easy $z$ dominant). As illustrated, the strain induced modulation in the anisotropy enables a drastic shift in the resonant frequencies.  For instance,  $\Delta f_{r,y}$ and $\Delta f_{r,z}$ are estimated to be around $493$ GHz and $296$ GHz, respectively, with the $E$ field varying from 0 to $-10$ kV/cm. These correspond to  $86$\% and $51$\% of the central frequencies at zero bias, which is quite sizable.  The modulation range can be further extended by considering the $E$ field of opposite polarity (i.e., $E > 0$).  One caution is that the validity of the model may become questionable in the region near the $x$$\leftrightarrow$$z$ transition point ($ E \sim +4.1 $ kV/cm) due to the absence of the "dominant" easy axis. Figure 3(b) examines $\mu_{yy}^{[x]}$ and $\mu_{zz}^{[x]}$ for small electric field strengths: i.e., $E = 0$, $\pm$0.5 kV/cm, and $\pm$1 kV/cm.  Compared to NiO, Cr$_2$O$_3$ can apparently provide a much wider range of eigen-frequency control so long as a similar level of the field can be applied.

To examine the issue of the linewidth, the dissipation term needs to be brought back into consideration. This makes the susceptibility  a complex quantity as discussed in Eqs.\ (\ref{chi_yy_c}) and (\ref{chi_zz_c}).
The plots in Figs.\ 4(a) and  4(b) provide the real and imaginary parts of $\chi_{yy}^{[x]}$ calculated with $\lambda$ ranging from $10^{-5}$ to $10^{-3}$ in NiO.  The choice of $\lambda$ clearly affects both results, making them continuous and finite.  In general, a  larger value of $\lambda$ (i.e., stronger dissipation) tends to wash out the characteristic resonance features (e.g., singularities).  The qualitatively different behaviors at/near the resonant frequency in these quantities (i.e., the real and imaginary parts of $\chi_{yy}^{[x]}$) can be readily understood from the quadratic vs.\ linear dependence on $\lambda$ in the second term of the corresponding numerator in Eq.\ (7).  The separation $\delta f$ in frequency between the peak and the valley in the real part (astride the former singularity) can be reasonably considered as the linewidth.  Alternatively, the full width at the half maximum height of the nicely peaked imaginary part can give a similar result.
An estimate of $\delta f$ shown in Fig.\ 4(c) clearly indicates that the linewidth of the EM signal could be rather insignificant compared to the predicted range of frequency modulation (i.e., sub to a few MHz vs.\ 10's to 100's of GHz). $\delta f$ appears to have a linear dependence on $\lambda$ with the coefficient of determination $R^2=99.99\%$ for both materials.

In addition to the well-anticipated shift in the frequencies, electrical control of the anisotropy profile can enable us to engineer other characteristic properties of interest.  For instance, the ellipticity of the EM waves at a given frequency can be modulated actively.
Assuming a incident signal circularly polarized on the $y$-$z$ plane, the same polarization is maintained only when $\mu^{[x]}_{yy}=\mu^{[x]}_{zz}$.
For convenience, the "ellipticity" of the EM wave may be quantified, similarly to the definition of eccentricity in geometry, as:
\begin{equation}
    e=\sqrt{\frac{b_l^2- b_s^2}{b_l^2}} ,
\end{equation}
where $b_l$ and $b_s$ refer to the larger  and  the smaller of $b_y$ and $b_z$ (where $\textbf{b} =\overleftrightarrow{\mu} \textbf{h}$). Accordingly, $e=1$ ($e=0$) corresponds to linear (circular) polarization, respectively, with $0\leq e\leq 1$ by definition.

Figures 5(a) and 5(b) show the results obtained in NiO and Cr$_2$O$_3$ as a function of the $E$ field for different incoming signal frequencies.  In the calculation, $\lambda = 10^{-4}$ is assumed to avoid the singularity and the imaginary part of the permeability is ignored for simplicity.   The cases analyzed are selected by considering the characteristic resonance frequency ranges.  More specifically, three frequencies examined in Fig.\ 5(a) (i.e., $75$ GHz, $225$ GHz, and $150$ GHz) represent the ranges of $f_{r,y}$, $f_{r,z}$, and the gapped region in NiO, respectively, as can be seen from Fig.\ 2(a).   Those for Cr$_2$O$_3$ in Fig.\ 5(b) are chosen similarly.  An interesting feature evident from the figures, beside a large variation in $e$, is that the $e=1$ peaks appear as a pair separated by a steep and narrow $e=0$  valley.  For a more detailed analysis, let us consider the case of $f=225$ GHz in NiO as an example.  Since 225 GHz is within the range of $f_{r,y}$, the $\mu_{yy}^{[x]}$ value at this frequency can change widely with the applied bias field while $\mu_{zz}^{[x]}$ stays essentially at 1.  When $\mu_{yy}^{[x]}$ hits $-1$ at the corresponding $E$, the ellipticity $e$ becomes zero since $b_y = b_z$.  However, a small deviation from this condition on either side of $E$ can lead $\mu_{yy}^{[x]} $ close to zero or to a much larger negative value. As such, $e$ increases  rapidly toward $1$, forming a pair of peaks near the $e=0$ valley.  This development occurs within a narrow window of $E$ (thus, frequencies).  Away from the peaks, $e$ starts to drop toward zero as $\mu_{yy}^{[x]}$ gradually converges to 1.  Another interesting observation is that the larger of the two in $b_y$ and $b_z$ [i.e., $b_l$ as defined in Eq.\ (17)] switches across the $e=0$ condition as denoted by the solid ($b_y > b_z$) vs.\ dashed ($b_z > b_y$) lines in Fig.\ 5(a).  They are characterized by the different principal axes of polarization ellipse and thus have different polarizations even though an identical $e$ value is assigned.  The result for $f=75$ GHz  also exhibits the double-peak feature and can be explained likewise.  By contrast, the case with $f=150$ GHz shows little dependence on $E$ since this frequency is away from the ranges of both $f_{r,y}$ and $f_{r,z}$.  Note that the region for $E > + 3.3$ kV/cm is not considered in the analysis due to the concern on the validity of the model as discussed earlier in relation to Fig.\ 2.

The situation in Cr$_2$O$_3$ is a bit more complex.  When no bias is applied ($E=0$), the uniaxial easy axis along the $x$ direction naturally leads to circular polarization on the $y$-$z$ plane (i.e., $e=0$) at any frequency.  Away from $E=0$, the material anisotropy becomes biaxial and the features similar to NiO start to develop.  Of the three signal frequencies chosen, $f=900$ GHz crosses or comes close to the resonance frequency once as $E$ changes (more specifically, $f_{r,y}$), whereas $700$ GHz and $300$ GHz do this twice (once each with $f_{r,y}$ and $f_{r,z}$ and twice with $f_{r,y}$, respectively); see also Fig.\ 3(a), where $f_{r,x}$ in line 3 is not part of the consideration.  Accordingly, the corresponding number of peak pairs can be expected in $e$ for each case as shown in Fig.\ 5(b).  Since the peaks and valleys are very sharp and narrow, it may be challenging to practically access the large $e$ region in Cr$_2$O$_3$.
Nevertheless, a sizable variation in the polarization can be achieved in both NiO and Cr$_2$O$_3$, which may be used to encode information in the EM signal with electrical control.  If the thickness of the magnetic layer can be kept in the range of $\mu$m or smaller, the characteristic time for electrically induced polarization change (thus, encoding) can be in the nanoseconds or shorter with a correspondingly high baud rate.

To examine the actual impact of the electrically modulated magnetic susceptibility in the device environment, the $S$ parameters
are evaluated in a simple (unoptimized) two-port configuration such as that shown in Fig.\ 6(a). Along with the material parameters specified earlier, the dimensions of the AFM Cr$_2$O$_3$ layer are assumed to be $L=2$ mm and $d=1$ mm in length and width.  With the thickness $t$ of the AFM film typically much smaller than 1 mm (e.g., a few $\mu$m or even thinner),~\cite{Sun2012} the cutoff frequency $f_c$ is given by $\frac{c}{2d}$, where $c$ is the speed of light.
In addition, the permittivity  $\epsilon$ is set to $12$ or 12$\epsilon_{0}$ in SI units.
An input signal polarized linearly along the $y$ axis is assumed in order to examine the response over the entire bias range under consideration [see line 1
in Fig.\ 3(a)].~\cite{Yang2011, Domich1991}

The result plotted in Fig.\ 6(b) clearly illustrates a sharp drop in $S_{11}$ as well as a visible increase in $S_{12}$ ($= S_{21}$) near the singularity of the permeability ($\sim574$ GHz with $E=0$), indicating very low reflection of the incident wave (thus, a desired characteristic of a sensitive bandpass filter).  The characteristic frequency for this precipitous decrease in $S_{11}$  (and thus a major jump in $S_{12}$) can apparently be tuned electrically as illustrated in Fig.\ 6(c).  The one that is observed around $275$ GHz corresponds to $E\sim +3.1$ kV/cm or $+5$ kV/cm, while  the other near $1.07$ THz is with $E=-10$ kV/cm.
As both show the drops exceeding $-20$ dB, an efficient bandpass filter with a  wide dynamic range appears feasible  in certain dielectric AFMs.  Note that these calculations  are performed with $\lambda$ of $ 5 \times 10^{-4}$ (a value somewhat larger than that used above for more smoothed out curves around the resonant frequencies).
While the presence of additional features as shown makes it more challenging to quantify the line broadening in the current example [i.e., Fig.\ 6(b)], it is safe to say that the width is in a few MHz range [in rough agreement with Fig.\ 4(b)].  Accordingly, a large number of signals may be placed within the modulated frequency window without overlap.  When $\lambda$ is increased to $10^{-3}$, the linewidth is still expected to be only around 10 MHz or so.   As the
operating frequency is moved higher (i.e., a shorter wavelength), the AFM EM channel can be made even thinner and surface effects such as magnetoelectric control of magnetic anisotropy may be adopted instead of the strain based approaches.

\section{Summary}
Voltage-controlled AFMR and frequency-dependent permeability are theoretically analyzed in the AFM/piezoelectric heterostructures.
By utilizing the susceptibility tensor derived for an EM wave with an arbitrary incident direction in uniaxial and biaxial AFMs, the calculation clearly shows that sub-mm wave components with a broad tunable spectral range in the 10's to 100"s of GHz (i.e., the resonant frequency shift) may be achieved via electrical control of magnetic anisotropy based on the piezoelectric and magneto-elastic effects.  Likewise, the polarization (e.g., ellipticity) of the EM signal can be modulated by exploring the anisotropic nature of the strain induced modulation (e.g., the in-plane vs.\ cross-plane anisotropy). The dependence on the incident direction of the EM signal is also analyzed. Application to a sample  bandpass filter validates the large range of central frequency shift in the $S$ parameters (i.e., tunability) coupled
with a narrow linewidth nearly independent of resonant frequency change, potentially
offering an extensive operable bandwidth.  Further, the dynamic reconfiguration can be achieve rapidly with an intrinsic delay in the ns order (assuming that the AFM layer is a few $\mu$m thick).
While Cr$_2$O$_3$ and NiO are considered as prototypical examples, other dielectric AFMs with strong magnetostrictive coupling are expected to have similar functionalities.   Our results clearly indicate the possibility of efficient application to electrically tunable microwave devices in the technically challenging THz regime.


\begin{acknowledgments}
This work was supported, in part, by the US Army Research Office (W911NF-20-2-0166).
\end{acknowledgments}

\clearpage
\begin{figure}
\includegraphics[width=8cm]{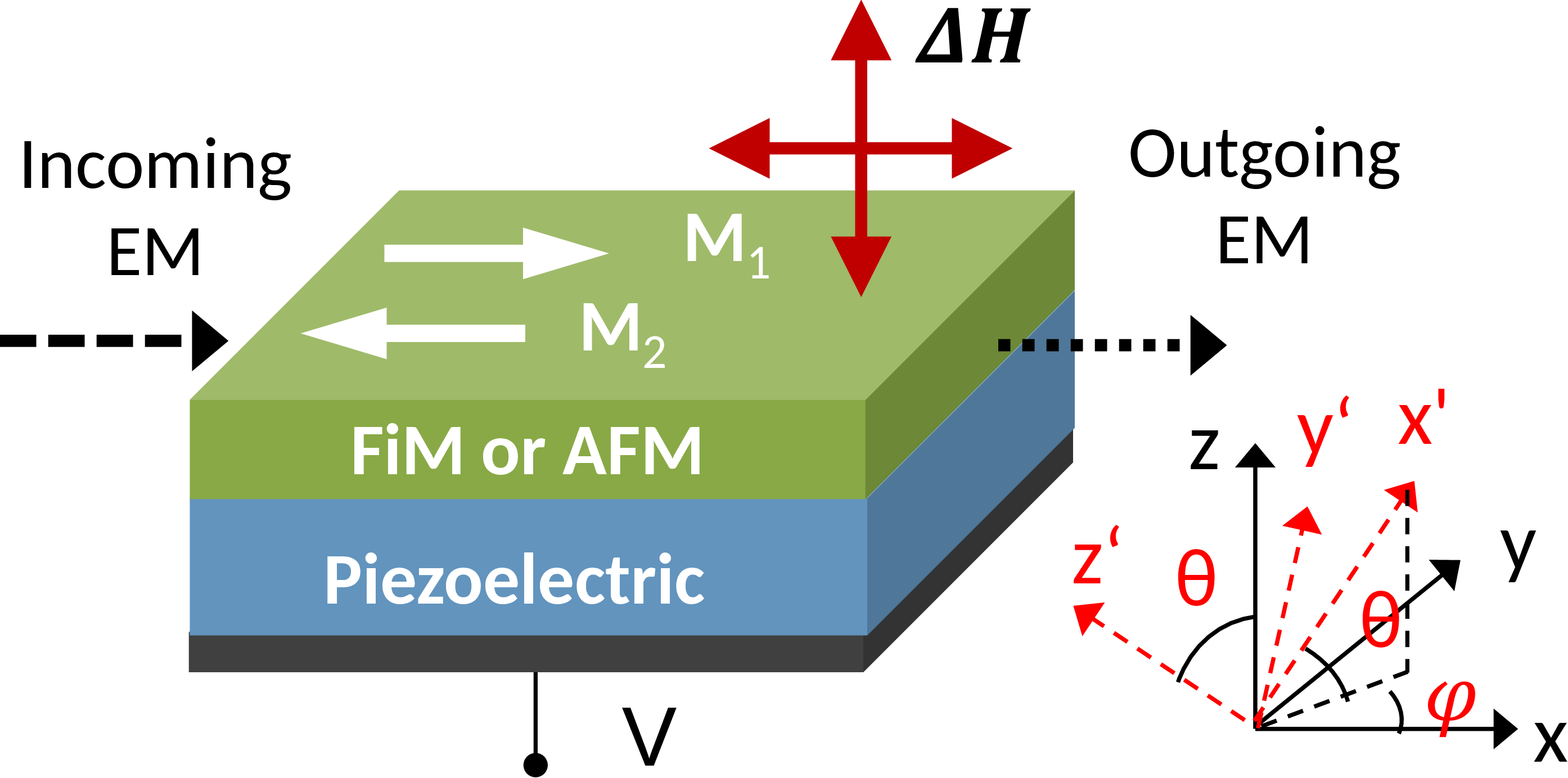}
\caption{Schematic of the  bilayer structure consisting of magnetic and  piezoelectric layers. The magnetic layer can be a uni-/bi-axial dielectric AFM or ferrimagnet (FiM)  with the in-plane and out-of-plane anisotropy axes.  The ground state of the magnetization is set initially along the ${x}$ axis. The EM wave is assumed incident in the same direction ($x$).  A bias applied to the piezoelectric layer can alter the magnetic anisotropy profile in the magnet via the magneto-elastic effect.  The Cartesian coordinates $x$,$y$,$z$ and $x^\prime$,$y^\prime$,$z^\prime$ are defined as shown along with the corresponding angles $\theta$, $\varphi$.}
\end{figure}
\clearpage

\begin{figure}
\includegraphics[width=8cm]{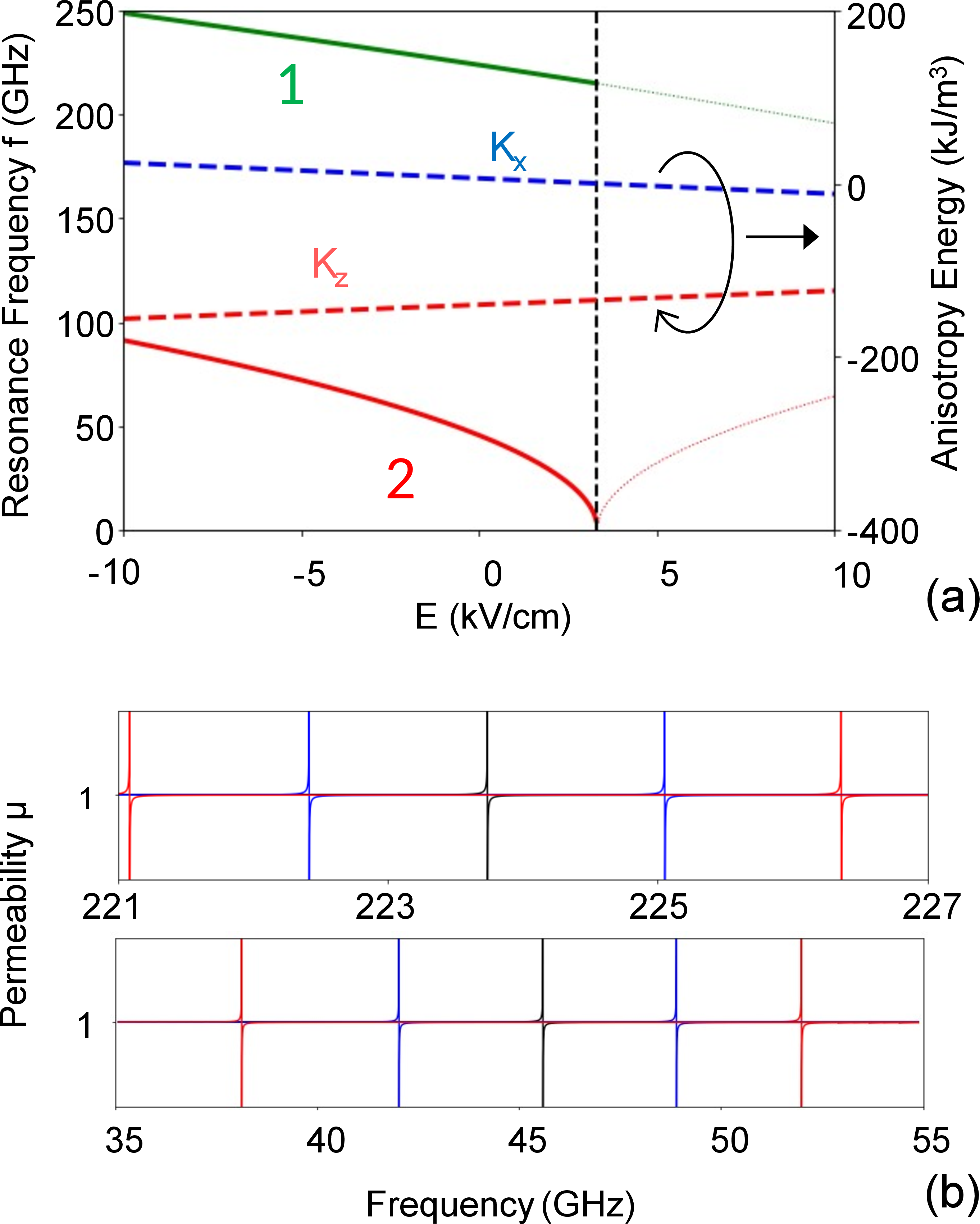}
\caption{EM response shift induced by strain in NiO. (a) Resonance frequencies and  magnetic anisotropy energies ($K_x$ and $K_z$) as a function of the applied electric field.  Lines $1$ and 2 show $f_{r,y}$ ($= \omega_{r,y}/2\pi$) and $f_{r,z}$ ($=\omega_{r,z}/2\pi$), respectively.  The region with $E > +3.3$ kV/cm (vertical line) is not considered in the analysis  due to the concern on the validity of the model.  (b) Permeability calculated with/without the modulation by the external electric field. The upper and lower panels show $\mu_{yy}^{[x]}$ and $\mu_{zz}^{[x]}$, respectively. Black lines: $E=0$; blue lines: $ E=\pm0.5$ kV/cm or $\Delta H_{A,x}=\mp50$ Oe; red lines: $ E=\pm1$ kV/cm or $\Delta H_{A,x}=\mp100$ Oe.  }
\end{figure}
\clearpage

\begin{figure}
\includegraphics[width=8cm]{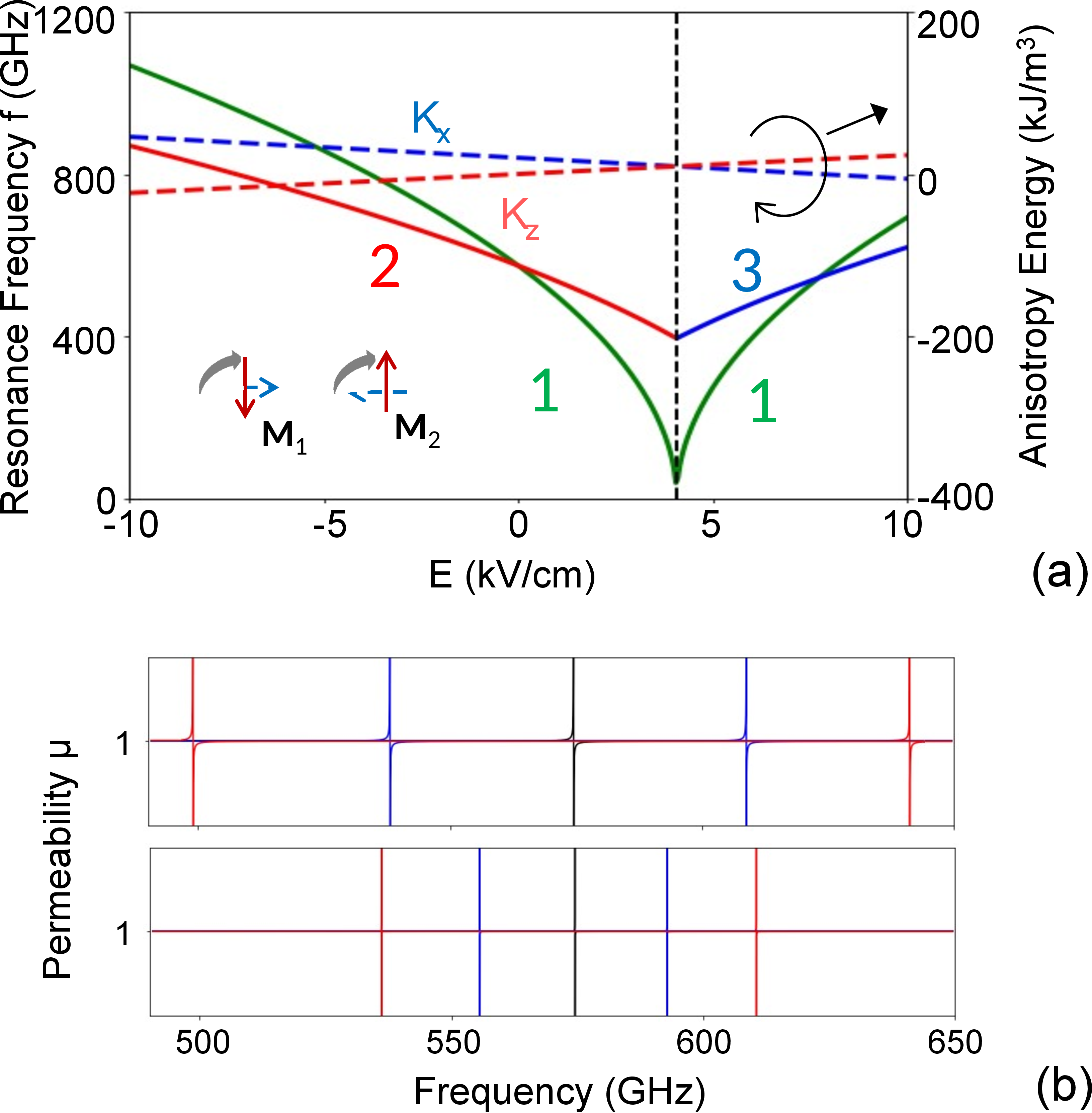}
\caption{EM response shift induced by strain in Cr$_2$O$_3$. (a) Resonance frequencies and  magnetic anisotropy energies ($K_x$ and $K_z$) as a function of the applied electric field.  Lines $1$, $2$, and $3$ are for $f_{r,y}$, $f_{r,z}$, and $f_{r,x}$, respectively. The inset shows the 90$^\circ$ magnetization rotation, from the in-plane to the out-of-plane orientations that occurs  at $E \sim +4.1$ kV/cm. Note that $f_{r,z}$, and $f_{r,x}$ are defined only in the corresponding configuration (i.e., in-plane and cross-plane, respectively).  The validity of the model may be in question in the transition region (i.e., near the vertical line).  (b) Permeability calculated with/without the modulation by the external electric field. The upper and lower panels show $\mu_{yy}^{[x]}$ and $\mu_{zz}^{[x]}$, respectively. Black lines: $E=0$; blue lines: $E=\pm0.5$ kV/cm or $\Delta H_{A,x}=\mp135$ Oe; red lines: $ E=\pm1$ kV/cm or $\Delta H_{A,x}=\mp270$ Oe.  }
\end{figure}
\clearpage

\begin{figure}
\includegraphics[width=8cm]{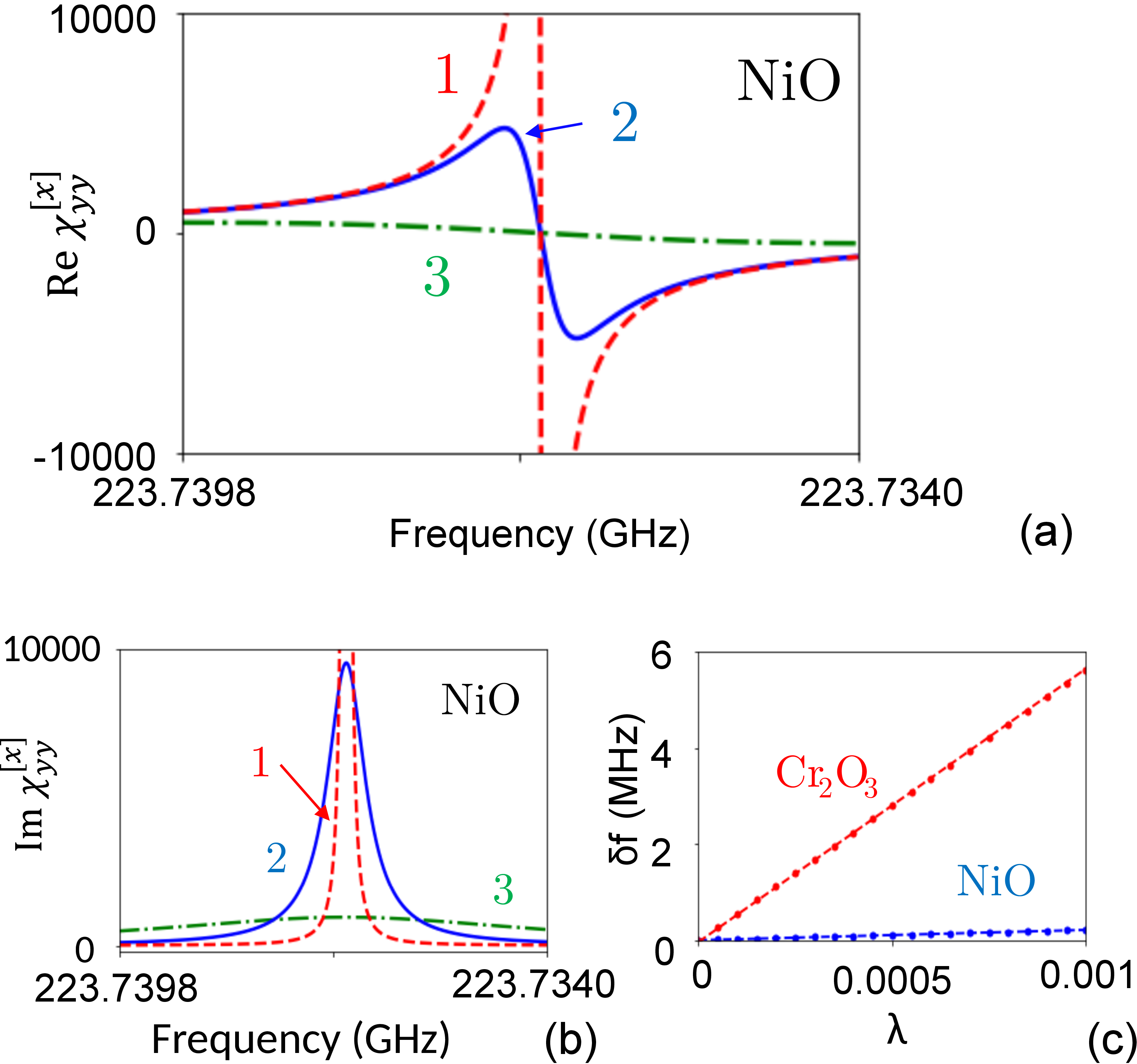}
\caption{ (a) Real and (b) imaginary parts of the susceptibility $\chi_{yy}^{[x]}$ in NiO as a function of frequency with $\lambda=10^{-5}$ (line 1), $10^{-4}$ (line 2), and $10^{-3}$ (line 3).  (c) $\delta f$ vs.\ $\lambda$ for NiO and Cr$_2$O$_3$ with linear regression lines.}
\end{figure}
\clearpage

\begin{figure}
\includegraphics[width=7cm]{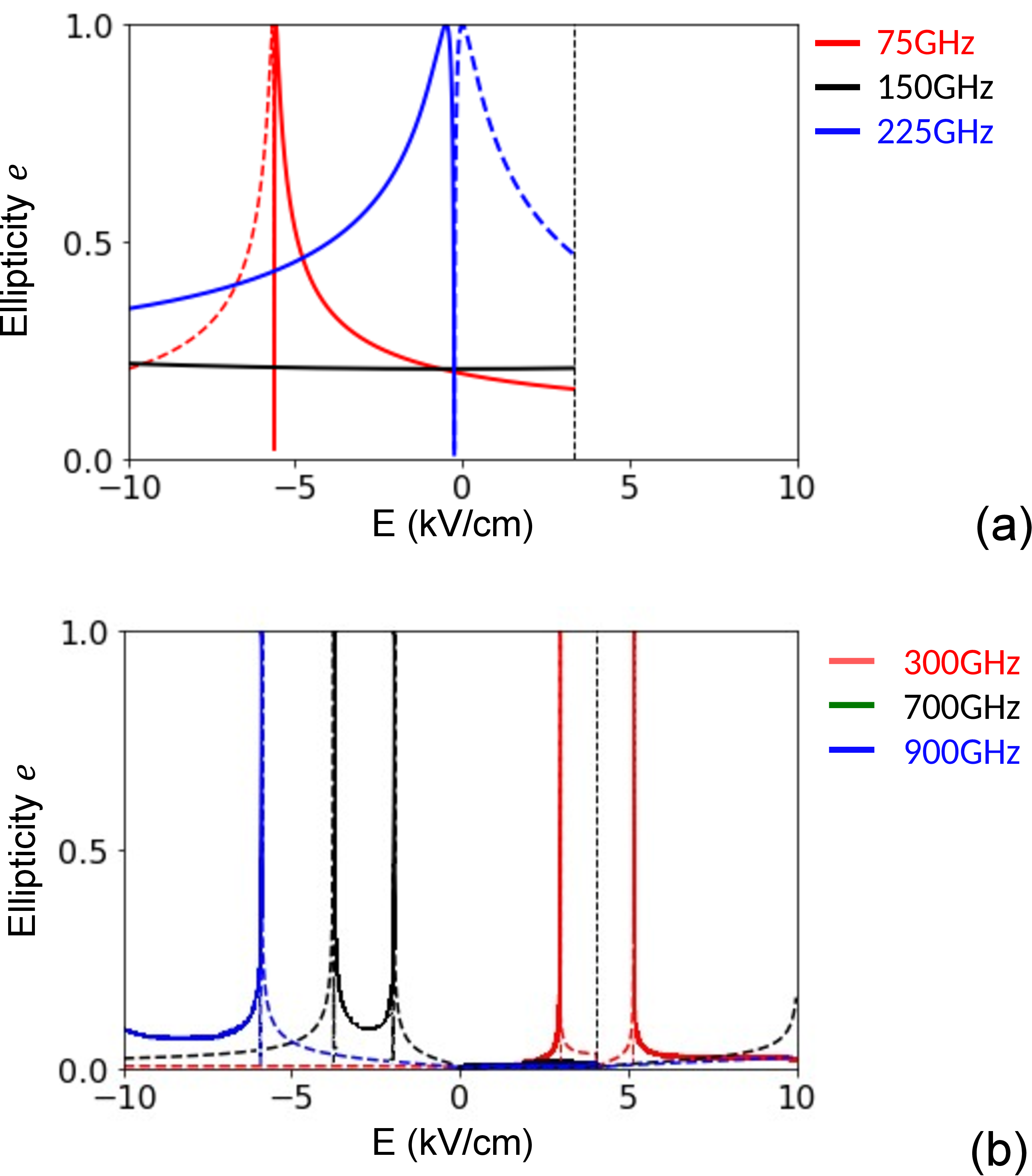}
\caption{Modulation of ellipticity in (a) NiO and (b) Cr$_2$O$_3$ as a function of the applied electric field.  The EM wave incident along $x$ is assumed to be circularly polarized with a frequency $f$.  The solid and dashed lines represent the cases when the principal axis of the polarization ellipse is in the $y$ and $z$ directions, respectively.   In (a), the region with $E > +3.3$ kV/cm (vertical line) is not considered in the analysis due to the concern on the validity of the model.  The vertical line in (b) denotes the condition at which the primary magnetic easy axis switches between the $x$ and $z$ directions. }
\end{figure}
\clearpage

\begin{figure}
\includegraphics[width= 16 cm]{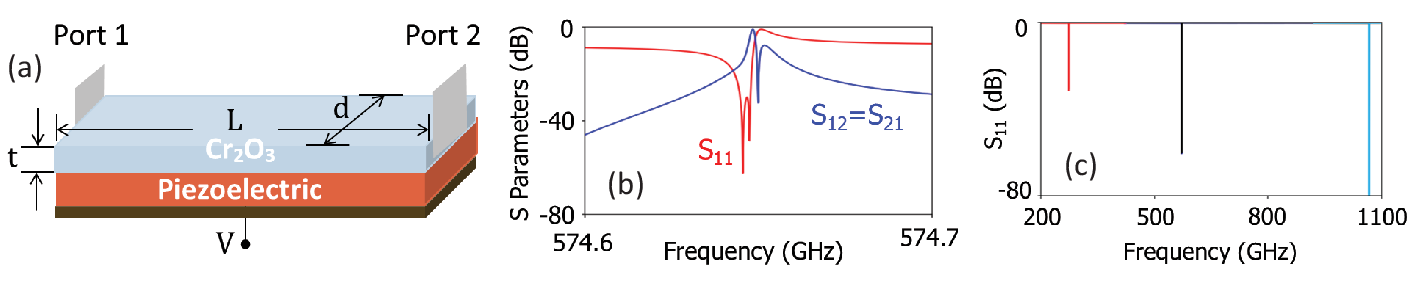}
\caption{(a) Schematic of a two-port configuration used in the $S$ parameter calculation. (b) $S_{11}$ and $S_{12} $($=S_{21}$) near the resonance frequency at $E=0$. (c) $S_{11}$ obtained at $E = +3.1$, $0$, and $-10$ kV/cm, respectively (from left to right).  An input signal polarized linearly along the $y$ axis is assumed with $\lambda = 5 \times 10^{-4}$. }
\end{figure}
\clearpage
\end{document}